\documentstyle[prc,aps,epsfig,multicol]{revtex}
\draft
\begin{document}
\title{Partonic effects on the elliptic flow at relativistic heavy ion 
collisions}
\author{Zi-wei Lin and C. M. Ko}
\address{Cyclotron Institute and Physics Department,
Texas A\&M University, College Station, Texas 77843-3366}
\maketitle

\begin{abstract}
The elliptic flow in heavy ion collisions at RHIC is studied in 
a multiphase transport model. By converting the strings in the high 
energy density regions into partons, we find that the final elliptic 
flow is sensitive to the parton scattering cross section. 
To reproduce the large elliptic flow observed in Au+Au collisions 
at $\sqrt s=130A$ GeV requires a parton scattering cross section of 
about 6 mb. We also study the dependence of the elliptic flow on the particle
multiplicity, transverse momentum, and particle mass.

\end{abstract}
\vspace{0.5cm}
\vspace{0.5cm}

\begin{multicols}{2}

\section{Introduction}

Elliptic flow in heavy ion collisions measures the asymmetry of particle
momentum distributions in the plane perpendicular to the beam direction. 
It results from the initial spatial asymmetry in non-central collisions
\cite{Barrette:1994xr,Appelshauser:1998dg}. 
Theoretical studies have shown that the elliptic 
flow is sensitive to the properties of the hot dense matter formed during 
the initial stage of heavy ion collisions 
\cite{Ollitrault:1992bk,Rqmd,Danielewicz:1998vz,Zheng:1999gt}.

Recently, elliptic flow has been measured at RHIC in Au+Au collisions at 
$\sqrt s=130A$ GeV. A large elliptic
flow of all charged particles near midrapidity was reported by the STAR 
collaboration \cite{Ackermann:2001tr}. 
Also, the transverse momentum dependence of the
elliptic flow of all charged particles was measured by both the STAR
\cite{Ackermann:2001tr} and the PHENIX \cite{Lacey:2001va} 
collaboration, while those
of charged pions, kaons, protons and anti-protons were measured by 
the STAR collaboration \cite{Snellings:2001nf,Adler:2001nb}. 
The experimental results show 
that the elliptic flow first increases with particle transverse momentum 
and then levels off. The dependences of elliptic 
flow on both the charged particle multiplicity 
\cite{Ackermann:2001tr,Lacey:2001va,Snellings:2001nf,Phobos:v2} 
and the particle pseudorapidity 
\cite{Phobos:v2} have also been measured. 

To understand these experimental results, many theoretical approaches 
have been used. These include semi-analytic models 
\cite{Heiselberg:1999mf}, models with parton energy loss 
\cite{Gyulassy:2001gk,Wang:2001fq}, 
hydrodynamic models \cite{Ollitrault:1992bk,Huovinen:2001cy,Huovinen:2001wn},  
transport models \cite{Rqmd,Zabrodin:2001rz}, and the hybrid model
which combines hydrodynamic and hadronic transport models \cite{Teaney:v2}. 
Among these studies, hydrodynamic models usually give the largest 
elliptic flow and an almost linear increase in its value  
with the particle transverse momentum below 1.5 GeV$/c$.  
In the hybrid model of combining the hydrodynamic model with 
the RQMD transport model \cite{Sorge:1995dp} and choosing certain effective 
equation of state, it is possible to 
obtain an elliptic flow that is comparable to the measured ones in heavy 
ion collisions at both SPS and RHIC energies \cite{Teaney:v2}. 
In transport models including only the parton cascade, 
the elliptic flow has been shown to be
sensitive to the parton scattering cross section, and a 
large value can be obtained with a large cross section 
\cite{Zhang:1999rs,Molnar:v2}.  On the other hand, 
transport models based on hadronic and/or string degrees of freedom 
in general give a smaller elliptic flow \cite{Snellings:2001nf} 
than that observed at RHIC. 
Including multi-Pomeron exchanges and hard gluon-gluon 
scatterings can, however, yield a large elliptic flow \cite{Zabrodin:2001rz}.

In this paper, we study the elliptic flow in Au+Au collisions 
at RHIC using a multiphase transport model (AMPT) that includes
both initial partonic and final hadronic interactions as well as 
the transition between these two phases of matter 
\cite{Zhang:2000bd,Lin:2001cx,Lin:2001yd}. 
In particular, we study the effect due to partons converted 
from the initial strings in high energy density regions.

\section{Conversion of strings to partons}

In most transport models for heavy ion collisions, such as the ART model 
\cite{Li:1995pr}, the RQMD model \cite{Sorge:1995dp}, 
and the UrQMD model \cite{Bass:1998ca}, 
initial primary collisions produce either hadrons or strings which 
later fragment to hadrons.  In the HIJING model \cite{Hijing},
on the other hand, initial primary collisions also produce minijet partons
which later enter into the string configurations and fragment to hadrons. 
These minijet partons are partons produced 
from initial hard collisions, i.e., from perturbative QCD processes 
involving a minimum transverse momentum transfer $p_0$, which is 
chosen to be 2 GeV$/c$ in the HIJING model to reproduce 
the $pp$ and $p\bar p$ data \cite{Hijing}. 

In the AMPT model \cite{Zhang:2000bd,Lin:2001cx,Lin:2001yd}, 
which takes initial conditions 
from the HIJING model, minijet partons first undergo scatterings before 
fragmenting into hadrons.  
Since the number of hard collisions in an $A+A$ collision  
roughly scales as $A^{4/3}$ and grows faster with colliding energy,
while the number of strings roughly scales as $A$, minijets become more 
important in heavy ion collisions at higher energies. 
However, for central Au+Au collisions even at $200A$ GeV minijet partons 
account for only about 1/3 of the total produced transverse energy, so 
the effect of parton scattering on the final particle multiplicities 
and spectra is quite small \cite{Lin:2001cx,Lin:2001yd}.  

The above picture of coexisting partons and strings during the initial
stage of high energy heavy ion collisions is questionable when the  
energy density is much higher than the critical density for the QCD phase 
transition. In this case, the strings are expected to 
melt into partonic degrees of freedom. 
Both the transport model \cite{Zhang:2000nc} 
and the high density QCD approach \cite{Kharzeev:2001ph} predict that 
the initial energy density of produced matter 
in central Au+Au collisions at RHIC is more than an order of 
magnitude higher than the critical energy density ($\sim 1$ GeV/fm$^3$).
Keeping strings in the high energy density region thus underestimates the
partonic effects in these collisions.

To model the string melting in high energy density regions to partons, 
we extend the AMPT model in the following way. After using the HIJING model 
(with jet quenching turned off) to produce the initial conditions, 
we first let strings fragment to hadrons using the LUND fragmentation 
\cite{Lund,Lin:2001cx} built in the PYTHIA routine 
\cite{Sjostrand:1994yb} 
and then convert these hadrons to partons according 
to their flavor and spin structure. In particular, 
a meson is converted to a quark and an anti-quark, while 
a baryon is converted to three quarks, 
and an anti-baryon is converted to three anti-quarks, 
where quark masses are taken as the same as in the PYTHIA program 
\cite{Sjostrand:1994yb}, e.g. $m_u=5.6$ MeV, $m_d=9.9$ MeV, $m_s=199$ MeV.
We further assume that 
quarks are produced isotropically in the rest frame of the hadron 
and start to interact only after a proper formation time $\tau_0^P$. 
In converting hadrons to partons, hadrons are not assigned 
a formation time as they are considered as an intermediate step
in modeling the melting of strings to partons in an environment 
of high energy density. Based on the expectation that hard partons 
(e.g. those described by perturbative QCD) 
are produced early while soft partons are 
produced late in the process, we take 
$\tau_0^P=1/Q$ for the parton proper formation time, 
where $Q$ is a scale related to the parton transverse momenta. 
Since partons produced from a string through the same intermediate
hadron should have the same formation time, the transverse mass of the  
hadron is thus used for $Q$ in determining the parton formation time.

The scatterings among these
quarks are treated using the parton cascade ZPC with 
the following universal cross section \cite{Zhang:1998ej}:
\begin{eqnarray}
\frac{d\sigma_p}{dt}=\frac{9\pi\alpha_s^2}{2} 
\left (1+{{\mu^2} \over s} \right ) \frac{1}{(t-\mu^2)^2}.
\end{eqnarray}
In the above,  
the strong coupling constant $\alpha_s$ is taken to be 0.47, $s$ and $t$ 
represent the standard Mandelstam variables for the two-parton elastic 
scattering process, and the effective screening mass $\mu$ depends
on the temperature and density of the partonic matter. In the present
study, we shall take $\mu$ as a parameter to obtain the desired 
total cross section and the corresponding angular distribution.

A parton can hadronize after it stops interacting, i.e., 
after it will no longer collide with other partons. 
We model the hadronization by combining the nearest two partons into a meson 
and three partons into a baryon (or an anti-baryon). 
As partons freeze out at different times and parton coalescence occurs 
at different times, the hadronization is treated locally. 
Since combinations of partons form a continuous invariant-mass spectrum  
but not a discreet one, it is in general impossible to conserve 4-momentum 
when several partons are combined into a hadron. 
In our current treatment, we choose to conserve 
the 3-momentum and determine the hadron species according to the flavor and 
invariant mass of coalescing partons. 
E.g., if a $\bar u$ and a $d$ quark 
coalesce, a $\pi^-$ will be formed if the 2-quark invariant mass is closer to 
the $\pi^-$ mass, or a $\rho^-$ will be formed if the 2-quark invariant 
mass is closer to the central value of the $\rho$ mass \cite{energy}. 
All SU(3) mesons and baryons listed in the HIJING program \cite{Hijing}
are included for coalescence  
except $\eta^\prime, \Sigma^*$ and $\Xi^*$, which are not present in 
our hadronic transport model, and $K_S^0, K_L^0$ states.
The resulting hadrons are then given an additional formation time of 
$\tau_0^H=0.7$ fm$/c$ in their rest frame and imported to the
ART hadronic transport model to take into account their rescatterings
\cite{Zhang:2000bd,Lin:2001cx,Lin:2001yd,Li:1995pr}. 

\section{Results on elliptic flow}

\subsection{Time evolution}

\begin{figure}[h]
\centerline{\epsfig{file=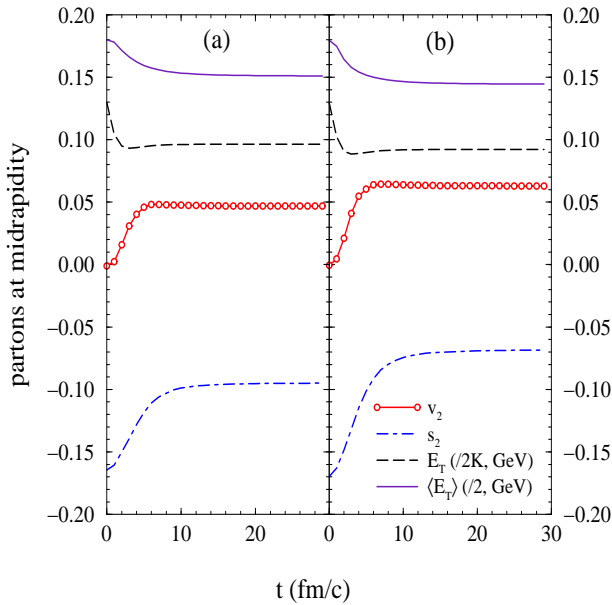,width=3.2in,height=3.2in,angle=0}}
\vspace{0.5cm}
\caption{Time evolutions of the parton elliptic flow, spatial anisotropy, 
transverse energy, and average transverse energy 
per parton at midrapidity in Au+Au collisions at center-of-mass energy
$130A$ GeV and impact parameter $b=8$ fm for parton scattering cross 
sections $\sigma_p=3$ mb (a) and 6 mb (b).} 
\label{figv2p}
\end{figure}

We have used the above extended AMPT model to study the elliptic flow
in Au+Au collisions at RHIC energies. 
The elliptic flow here is defined as 
\begin{equation}
v_2 =\left \langle {{p_x^2-p_y^2} \over {p_x^2+p_y^2}} \right \rangle, 
\end{equation}
where the average is performed over all particles. 
For center-of-mass energy 
$\sqrt s=130A$ GeV and impact parameter $b=8$ fm, we show in 
Figs. \ref{figv2p}(a) and \ref{figv2p}(b) the time evolution of the elliptic 
flow of all partons, i.e., regardless of their formation and freeze-out times, 
at midrapidity with parton scattering cross sections $\sigma_p=3$ and 6 mb, 
respectively. 
We note that partons still inside strings 
and partons which have frozen out are also included in Fig.~1. 
For both values of parton cross sections  
the elliptic flow is seen to develop mostly within the first 5 fm$/c$, 
and both the rate of increase and the final parton $v_2$ are larger for 
the larger parton cross section. Also shown in these figures
are the time evolutions of the parton transverse
energy $E_T$ (scaled down by a factor of 2000), 
the average transverse energy per parton $\langle E_T \rangle$ 
(scaled down by 2), and the second moment of the 
spatial asymmetry determined from the parton positions at their previous 
interaction points, i.e., $s_2=\langle (x^2-y^2)/(x^2+y^2)\rangle$. 
It is seen that for both parton cross sections 
the absolute value of $s_2$ decreases with time, 
and its final value is closer to zero in the case of $\sigma_p=6$ mb. 
The final saturation of parton $v_2$ in the partonic phase 
is due to the lack of scatterings in the limit of small $\sigma_p$ 
and the vanishing spatial anisotropy in the limit of large $\sigma_p$.
We also see that the transverse energy at midrapidity decreases by $\sim$ 30\% 
within a couple of fm$/c$. 
Since the decrease of parton transverse energy 
is faster than that for the parton number at midrapidity,  
the average transverse energy per parton $\langle E_T \rangle$ also
decreases.  The final saturation of $\langle E_T \rangle$ at
$\sim$ 20\% below the initial value reflects the equilibration
between the longitudinal and transverse momenta of partons 
\cite{Gyulassy:1997ib,Dumitru:2000up}. 

\begin{figure}[h]
\centerline{\epsfig{file=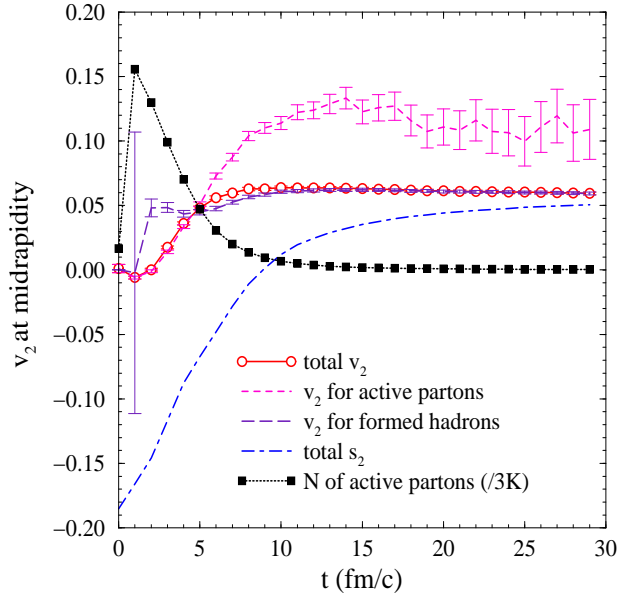,width=3.2in,height=3.2in,angle=0}}
\caption{Time evolutions of elliptic flow, spatial anisotropy,
and the number of active partons 
at midrapidity in Au+Au collisions at center-of-mass energy of $130A$ 
GeV and impact parameter $b=8$ fm for a parton cross section $\sigma_p=6$ mb.} 
\label{figv2ph}
\end{figure}

In Fig. \ref{figv2ph}, we show the time evolution of $v_2$ for active partons,
i.e., partons that have not stopped scatterings, $v_2$ for 
formed hadrons, the total $v_2$ and $s_2$ including both active partons and 
formed hadrons, and the number of active partons at midrapidity for 
collisions at an impact parameter $b=8$ fm and a parton cross 
section $\sigma_p=6$ mb. 
While the initial numbers of active partons are near zero 
because most partons are not yet formed and thus not yet active,
it is seen that the number of active partons 
peaks within the first fm$/c$ and decreases by a factor of 2 at about 4 fm$/c$ 
due to the hadronization. 
The total $v_2$ is thus dominated by active partons at the early stage and by 
formed hadrons during the later stage of heavy ion collisions. 
Although the active partons have a large elliptic flow at later times, 
their number is too small to affect the total elliptic flow. 

\subsection{Impact parameter dependence}

We have also studied the impact parameter dependence of elliptic flow 
in Au+Au collisions at center-of-mass energies of $130A$ GeV.
To compare with the centrality dependence in the STAR data
\cite{Ackermann:2001tr}, we first divide the impact parameter
range $0\le b\le 13$ fm into six bins with equal bin-size except the
first bin, which is taken to be $0\le b\le 3$ fm. 
The impact parameter dependence is then converted to the dependence on
$N_{\rm ch}/N_{\rm max}$ by taking $N_{\rm ch}$ as the number 
of charged particles within the pseudorapidity range
$\eta \in (-0.75, 0.75)$ and $N_{\rm max}$ as its value at $b=0$ fm.
Furthermore, we include only charged particles within $\eta \in (-1.3, 1.3)$ 
and transverse momentum range $p_t \in (0.1, 2.1)$ GeV$/c$ in evaluating
the elliptic flow $v_2$ in order to compare more directly with the STAR data 
on the centrality dependence.

\begin{figure}[h]
\centerline{\epsfig{file=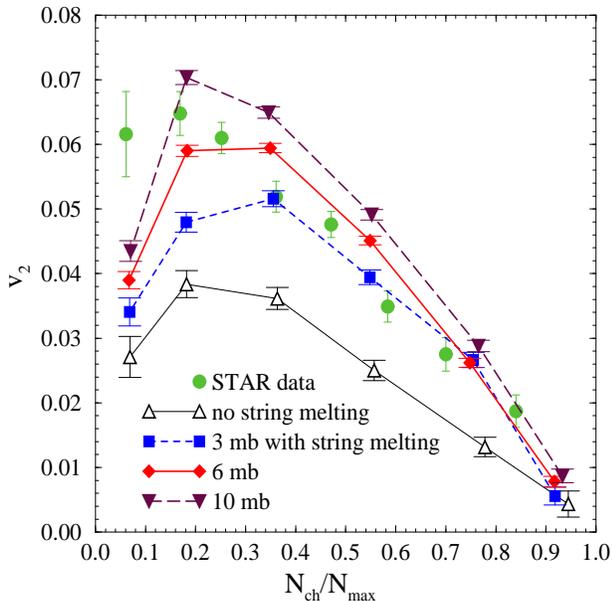,width=3.2in,height=3.2in,angle=0}}
\caption{Impact parameter dependence of elliptic flow at $130A$ GeV.
The data from the STAR collaboration
\protect\cite{Ackermann:2001tr} are shown by filled circles, while
the theoretical results for different partonic dynamics 
are given by curves.}
\label{figb}
\end{figure}

In Fig. \ref{figb}, we show the results for the elliptic flow of charged 
particles as a function of $N_{\rm ch}/N_{\rm max}$ 
for the scenarios of the default AMPT (without string melting) 
and the extended AMPT (with string melting).  
The error bars in our results represent only the statistical error 
in $v_2$ but not that in $N_{\rm ch}$. Although
all results show the qualitative features of the observed 
centrality dependence of $v_2$ \cite{Ackermann:2001tr}, the shape and 
magnitude of $v_2$ depend sensitively on the partonic dynamics 
\cite{Voloshin:2000gs}. 
Without converting the initial strings into partons, the default AMPT 
model gives the smallest elliptic flow. Allowing the melting of strings 
to partons, the elliptic flow for a larger parton cross section 
not only is higher than that for a smaller partonic cross section, 
but also peaks at a lower value of $N_{\rm ch}/N_{\rm max}$. 
Of the three parton cross sections,
the results for $\sigma_p=6$ mb appear to be more consistent with 
the observed centrality dependence of the elliptic flow. 

We note that the results from the default AMPT model are insensitive to 
the parton scattering cross section, 
which is taken to be $3$ mb in the scenario with strings.
This is mainly due to the small fraction of energy that is 
carried by minijet partons and the lack of transverse collective
motion of the strings. As a result, 
the elliptic flow is significantly reduced after minijet partons combine with
strings and fragment to hadrons. However, with strings converting to partons, 
the initial energy originally stored in the strings also
contributes to the parton dynamics. 
This leads to a larger elliptic flow and its sensitivity 
to the parton cross section, and thus makes it possible to determine 
the strength of partonic interactions from the final elliptic flow.

\subsection{Transverse momentum dependence}

\begin{figure}[h]
\centerline{\epsfig{file=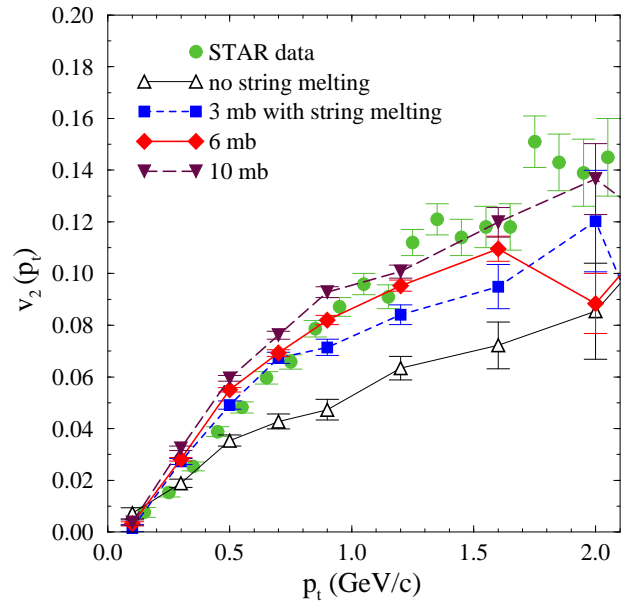,width=3.2in,height=3.2in,angle=0}}
\caption{Transverse momentum dependence of elliptic flow at $130A$ GeV.
Circles are the STAR data for minimum-bias Au+Au collisions 
\protect\cite{Ackermann:2001tr}, and curves represent the minimum-bias 
results for charged particles within $\eta \in (-1.3, 1.3)$
from the AMPT model.}
\label{figpt}
\end{figure}

Our results for the dependence of charged particle elliptic flow on
the transverse momentum, i.e., the differential elliptic flow $v_2(p_t)$,
are shown in Fig. \ref{figpt}. As observed in the experiment data, the 
differential elliptic flow first increases almost linearly with 
transverse momentum and then tends to level off at large transverse 
momenta. However, both the slope of initial increase and the 
transverse momentum at which deviation from a linear dependence 
appears are affected by the parton dynamics. 
The result from the default AMPT model 
\cite{Zhang:2000bd,Lin:2001cx,Lin:2001yd} 
(open triangles), which includes only minijets in the partonic phase, has
the smallest $v_2(p_t)$ at a given $p_t$ and shows a
departure from the linear dependence also at the smallest transverse 
momentum. Including partons from string melting in the high energy
density regions increases both the magnitude of $v_2(p_t)$ and 
the value of $p_t$ at which the linear dependence breaks down.

\begin{figure}[h]
\centerline{\epsfig{file=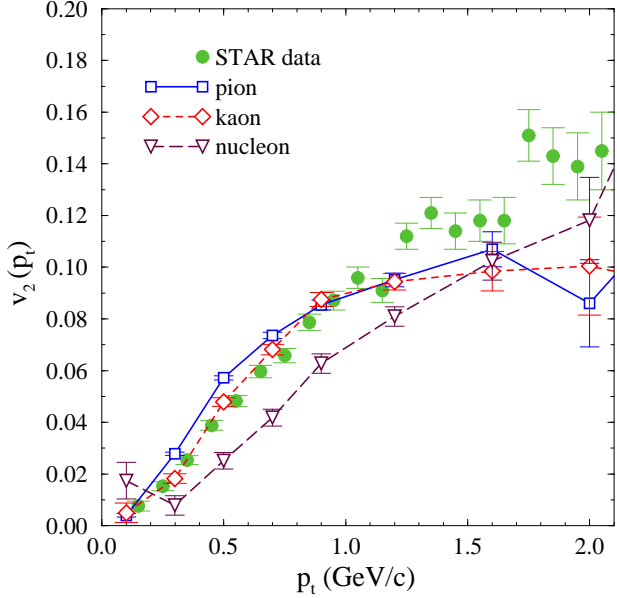,width=3.2in,height=3.2in,angle=0}}
\caption{Transverse momentum dependence of the elliptic flow 
for particles of different masses in the case of parton cross section 
$\sigma_p=6$ mb in Au+Au collisions at $130A$ GeV. 
Curves represent the minimum-bias results within $\eta \in (-1.3, 1.3)$ 
from the extended AMPT model.}
\label{figm}
\end{figure}

The differential elliptic flow $v_2(p_t)$ 
is expected to be different for different particles 
\cite{Heiselberg:1999es,Huovinen:2001cy}. 
Fig.~\ref{figm} shows our results for pions, kaons and nucleons 
from the extended AMPT model together with the STAR data for charged 
particles \cite{Ackermann:2001tr}. At low transverse momentum 
($p_t<1$ GeV$/c$), while the $v_2(p_t)$ of pions and kaons increases 
almost linearly with $p_t$, that of nucleons shows a stronger
dependence on $p_t$. Furthermore, particles with smaller masses 
are seen to have higher values of $v_2(p_t)$ at a given $p_t$. 
These features are qualitatively similar to those 
observed from the STAR data \cite{Adler:2001nb}
and also those obtained from the hydrodynamic models
\cite{Huovinen:2001cy,Huovinen:2001wn}. On the other hand, 
all particles seem to have similar values of elliptic flow,
within the errors of our calculations,
at a given $p_t$ above $1$ GeV/$c$.

To test the sensitivity of our results to different formation time parameters, 
we have calculated the elliptic flow at $b=8$ fm for 
the following three cases for comparison: 
i) parton formation time $\tau_0^P$ larger by a factor of 2, 
ii) $\tau_0^P$ smaller by a factor of 2,
iii) hadron formation time  $\tau_0^H$ set to 0.01 fm$/c$ 
instead of the default 0.7 fm$/c$.
Compared to our default results, the relative changes of final 
charged particle $v_2$ are found to be less than 10\% for $p_\perp <2$ GeV$/c$.

\subsection{Elliptic flow at $\sqrt s=200A$ GeV}

\begin{figure}[h]
\centerline{\epsfig{file=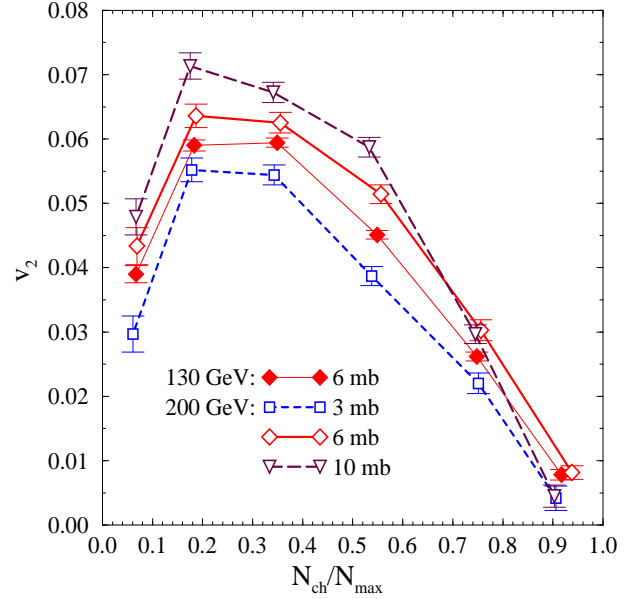,width=3.2in,height=3.2in,angle=0}}
\caption{Impact parameter dependence of elliptic flow at $200A$ GeV. 
Curves represent minimum-bias results for charged particles 
within $\eta \in (-1.3, 1.3)$ and $p_t \in (0.1, 2.1)$ GeV$/c$.}
\label{fig_b200}
\end{figure}

\begin{figure}[h]
\centerline{\epsfig{file=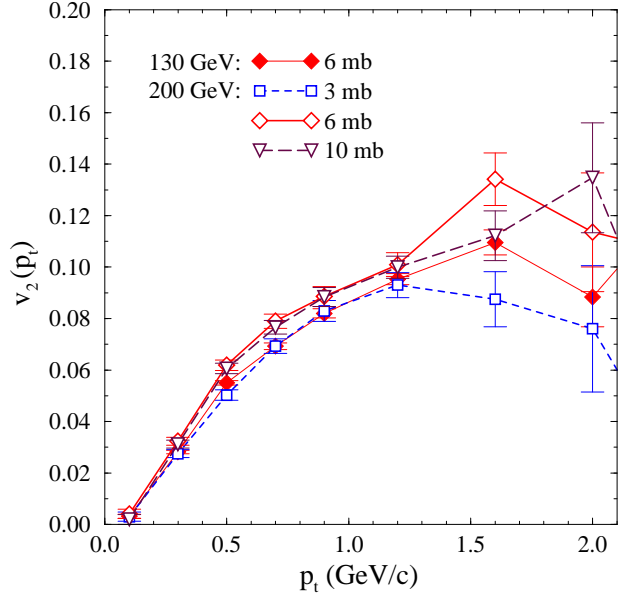,width=3.2in,height=3.2in,angle=0}}
\caption{Transverse momentum $p_t$ dependence of elliptic flow at $200A$ 
GeV. Curves represent minimum-bias results for charged particles 
within $\eta \in (-1.3, 1.3)$.}
\label{fig_pt200}
\end{figure}

We have also studied the elliptic flow in Au+Au collisions at the
maximum RHIC energy of $200A$ GeV using the extended AMPT model. 
In Figs. \ref{fig_b200} and \ref{fig_pt200}, we  
show the dependence of the charged particle elliptic flow on
impact parameter and transverse momentum, respectively,  
together with the results for $\sigma_p=6$ mb at $\sqrt s=130A$ GeV 
(curves with filled diamonds). 
Compared to Fig.~\ref{figb}, it is seen that 
for the same parton cross section the elliptic flow in less central events 
($N_{\rm ch}/N_{\rm max}<0.6$) increases only slightly (no more than 0.01
in magnitude) with the center-of-mass energy. As a result,
the sensitivity of the centrality dependence of elliptic flow to
the parton scattering cross section in heavy ion collisions at $\sqrt s=200A$
GeV is similar to that at $\sqrt s=130A$ GeV. 

For the differential elliptic flow, Fig. \ref{fig_pt200} shows that
for transverse momenta below about 1 GeV$/c$, it shows even less change with 
the center-of-mass energy than the impact parameter dependence of the
elliptic flow.  As to the dependence on the parton cross section, 
the differential elliptic flow seems to show a larger sensitivity 
at higher transverse momenta. However, to study this quantitatively requires 
much better statistics than we have obtained so far.
Also, for elliptic flow at high $p_t$,  
the effect of energy loss due to inelastic partonic processes, which have not
been included in the AMPT model, becomes important 
\cite{Gyulassy:2001gk,Wang:2001fq}. 
Furthermore, the parton coalescence model may be less
suitable than the independent parton fragmentation model
for modeling the hadronization dynamics. 

\section{Conclusions and discussions}

In this paper, we have studied the elliptic flow in heavy ion 
collisions at RHIC using a multiphase transport model that includes both
partonic and hadronic scatterings.  To take into account the effects
of string melting due to initial high energy density, 
we have introduced a schematic 
model to convert the strings produced in soft interactions
into partons. We find that the magnitude of the elliptic flow is  
sensitive to the scattering cross sections of these partons.
To reproduce the large elliptic flow observed in Au+Au collisions 
at $130A$ GeV at RHIC requires a parton scattering cross section of 
roughly 6 mb.  Similar to the findings from hydrodynamic models, 
the differential elliptic flow $v_2(p_t)$ for charged particles 
is found to increase almost linearly with $p_t$ at low transverse momentum. 
At high transverse momentum $v_2(p_t)$ deviates from 
a linear dependence and becomes more flat. Also, 
heavier particles have smaller $v_2(p_t)$ than lighter particles 
at a given low $p_t$, while they seem to have similar values of 
elliptic flow at higher $p_t$. 
We further find that the increase of the elliptic flow from $\sqrt s=130A$ 
to $200A$ GeV is quite modest in the centrality dependence 
and is even less in the dependence on transverse momenta below 2 GeV$/c$.
 
In the present study, we have adopted a simple approach in converting 
strings to partons and in the hadronization from partons to hadrons. 
To treat this more consistently, we need to consider the local 
parton and string densities and determine when strings are 
converted to partons and partons are combined to from hadrons.
In a probably more systematic approach, the initial energy may be 
separated into two parts with the hard part corresponding to minijet 
gluons and quarks, and the soft part corresponding to color fields. 
A parton cascade including a color mean field can then be
used to describe the subsequent dynamics.
Furthermore, both inelastic parton scatterings and 
particle subdivision \cite{Zhang:1998tj,Molnar:v2,Cheng:2001dz} 
have been neglected in the parton cascade. These may affect the
elliptic flow and need to be studied in the future.

\medskip

We appreciate useful discussions with 
U. Heinz, B.A. Li, T.W. Ludlam, M. Murray, J.Y. Ollitrault, 
Subrata Pal, D.H. Rischke, N. Xu and B. Zhang. 
This work was supported by the U.S. National 
Science Foundation under Grant Nos. PHY-9870038 and PHY-0098805, the Welch
Foundation under Grant No. A-1358, and the Texas Advanced Research
Program under Grant No. FY99-010366-0081.

\end{multicols}


\begin{thebibliography}{99}
\bibitem{energy}
We have checked that numerically the violation of total energy in any event 
is between -0.4\protect \% and 1.7\protect \% 
for 5000 minimum-bias Au+Au events at $130A$ GeV.
\end{thebibliography}
\end{document}